\documentclass[sn-nature,pdflatex]{sn-jnl}

\usepackage{graphicx}%
\usepackage{multirow}%
\usepackage{adjustbox}
\usepackage{amsmath,amssymb,amsfonts}%
\usepackage{amsthm}%
\usepackage{mathrsfs}%
\usepackage[title]{appendix}%
\usepackage{xcolor}%
\usepackage{textcomp}%
\usepackage{manyfoot}%
\usepackage{booktabs}%
\usepackage{algorithm}%
\usepackage{algorithmicx}%
\usepackage{algpseudocode}%
\usepackage{listings}%
\usepackage{caption}
\usepackage{fancyhdr}
\usepackage[normalem]{ulem}

\usepackage{array}
\usepackage{makecell}
\newcolumntype{x}[1]{>{\centering\arraybackslash}p{#1}}
\usepackage{tikz}

\begin{document}
\title[Active Learning of A Crystal Plasticity Flow Rule From Discrete Dislocation Dynamics Simulations]{Active Learning of A Crystal Plasticity Flow Rule From Discrete Dislocation Dynamics Simulations}

\author*[1]{\fnm{Nicholas} \sur{Julian}}\email{njulian@ucla.edu}
\author[2]{\fnm{Giacomo} \sur{Po}}\email{gpo@miami.edu}
\author[3]{\fnm{Enrique} \sur{Martinez}}\email{enrique@clemson.edu}
\author[1]{\fnm{Nithin} \sur{Mathew}}\email{mathewni@lanl.gov}
\author[1]{\fnm{Danny} \sur{Perez}}\email{danny\_perez@lanl.gov}
\affil*[1]{\orgname{Los Alamos National Laboratory}, \orgaddress{\state{NM}, \country{USA}}}
\affil[2]{\orgname{University of Miami}, \orgaddress{\state{FL}, \country{USA}}}
\affil[3]{\orgname{Clemson University}}


\abstract{
Continuum-scale material deformation models, such as crystal plasticity, can significantly enhance their predictive accuracy by incorporating input from lower-scale (i.e., mesoscale) models.
The procedure to generate and extract the relevant information is however typically complex and {\em ad hoc}, involving decision and intervention  by domain experts, leading to long development times. In this study, we develop a principled approach for calibration of continuum-scale models using lower scale information by representing a crystal plasticity flow rule
as a Gaussian process model. This representation
allows for efficient parameter space exploration, guided by the uncertainty embedded in the model through a process known as Bayesian optimization.
We demonstrate a semi-autonomous Bayesian optimization loop which instantiates discrete dislocation dynamics simulations whose initial
conditions are automatically chosen to optimize the uncertainty of a model crystal plasticity flow rule. 
Our self-guided computational pipeline efficiently generated a dataset and corresponding model whose error, uncertainty, and physical feature sensitivities were validated with comparison to an independent dataset four times larger,
demonstrating a valuable and efficient active learning implementation readily transferable to similar material systems.

}
\maketitle
\pagestyle{fancy}
\fancyfoot[L]{LA-UR-25-28614} 
\fancyhead{}
\section{Introduction}\label{sec1}
Dislocation-mediated plasticity is the primary mechanism for
inelastic 
deformation in crystalline materials. Computational methods of predicting deformation of single-  and poly-crystal materials are of great importance in a multitude of structural applications. 
A commonly used computational method is Crystal Plasticity (CP) which predicts continuum-scale deformation using constitutive representations of the various deformation phenomena occurring at lower scales using a so-called flow rule. The flow rule predicts single crystal strain rates along crystallographic slip systems in response to an applied load,
encapsulating complex motions and interactions of dislocations as a function of dislocation density and stress, see, e.g., Ref.\ \cite{ROTERS20101152}.

 The information required to customize the parameters of a flow rule to a specific material is of critical importance, spanning a large parameter space with many degrees of freedom. Methods for efficiently generating this data set from simulations of the underlying dynamics of dislocations are critical for expediting model improvements and adapting them to new material systems.

To date, flow rules in crystal plasticity have primarily, if not entirely, taken on forms based upon phenomenology or constitutive laws.
References~\cite{MA200791} and \cite{CHO2018138} illustrate and address the challenges in designing a flow rule which accurately represents the deformation dynamics of BCC structured materials using phenomenologically based formulae.
Some of these challenges may be addressed by inspecting simulations of BCC deformation dynamics at the atomic scale, and developing customized flow rule formulae to that data, as in \cite{LIM2015100} and \cite{BERTIN2023119336}. The drawback to this approach is that atomic simulations of sufficient size are so costly that opportunities to produce a wide enough variety of them to extract parameters of a flow rule are uncommon.

To avoid the high cost of atomic simulations, Discrete Dislocation Dynamics (DDD) simulations aim to predict crystalline deformation directly at the meso-scale by representing dislocations as simple line defects whose interactions are mediated by linear elasticity.
By enabling a significantly larger set of training data at sufficient size, DDD simulations can ~provide the physical insight required to propose corrections to legacy phenomenological relationships for plastic flow and dislocation multiplication, as well as to develop and fit new constitutive flow rules  \cite{AKHONDZADEH2020104152}.

Prior work along these lines used data generated from simulations or experiments whose parameters were chosen heuristically, producing an expert's best guess at an adequate representation of the parameter space of applied mechanical loads and dislocation configurations \cite{AKHONDZADEH2020104152,MA20043603,MA200791,LIM2015100,LEE1999625,BERTIN2023119336}.
Generating the corresponding data required human expertise and labor. To expedite and reduce the cost of generating the data required for fitting a model, it is highly desirable to automate the simulations through methods which optimize the parameter space exploration aided by an uncertainty metric directly embedded in the model.
In this work, we demonstrate a 
semi-autonomous computational pipeline which iteratively instantiates a sequence of DDD simulations, efficiently navigating the parameter space of a flow rule to build a data set sufficient to fit a crystal plasticity flow rule.

The manuscript is organized as follows. Section~\ref{sec2} discusses in detail our choices of: mesoscale model to be fitted (\ref{sec_modelSelection}), 
the parameterization of the model domain (\ref{sec_modelParameters}),
implementation of the chosen model form (\ref{sec_GPdetails}),
the method by which this model is used to identify simulation specifications to be instantiated (\ref{sec2Acqf}),
the constraints used to ensure physical and practical plausibility (\ref{sec2Con}),
the implementation of DDD used (\ref{sec2DDD}),
processing and preparation of the simulation measurements to be presented to the model (\ref{postprocessing}),
and the overall workflow and automation methods we implemented (\ref{sec2workflow}).
In Section~\ref{sec3}, we present examples of pertinent measurements from DDD (\ref{sec3DDD}),
and an assessment of the quality of the model as it evolved through the sequence of simulations which it guided (\ref{sec3model}).
Finally, we briefly summarize our achievements and conclude in Section~\ref{sec4}.
Several appendices are provided as supplementary information.
\section{Materials and Methods}\label{sec2}

\subsection{Model selection
}\label{sec_modelSelection}
Sequential exploration of a complex parameter space 
may
benefit
from insight from the model to guide its own improvement using a local measure of  uncertainty.
The high cost of simulations and thus likely restrictions on data set size must also be considered, as many candidate machine learning methods require a prohibitively large amount of data.
When choosing a model, its parameterization, and corresponding transformations to and from DDD simulation data 
in this context we prioritize optimizing the efficiency with which we navigate the high dimensional parameter space to gather data which is most effective at reducing predictive errors and at naturally encapsulating
the basic
underlying physical phenomena.
The data generated may be used in future work to fit models whose parameterization and fidelity may be tuned to optimize behavior in a CP context.

Gaussian processes (GP) have been broadly utilized in situations like these. 
Abstractly,
a GP is a collection of random variables, any finite subset of which follows a joint multivariate Gaussian distribution.
As a consequence, an instance of a
GP is fully specified by a mean function and a covariance (kernel) function
which may be fitted to data by optimizing their hyperparameters, 
allowing for flexible 
modeling of complex functions.
Additionally, GP models are well suited to be sequentially updated as new information becomes available through the application of Bayes' Rule since the initial state, called the prior, the prediction of the model, called the likelihood, and the resulting updated distribution, called the posterior, have easily obtainable formulae.
The variance of the posterior is then a natural quantification of the uncertainty of the model.
An effective active learning scheme is then achieved by instantiating simulations at points
for which the model's posterior distribution has a high variance, as chosen by a so-called acquisition function,
with the expectation that obtaining additional data at those points will reduce the said variance and thus the local model uncertainty.
Such a method of active learning that implements GP regression (GPR)
in this way is commonly known as Bayesian optimization (BO) \cite{Frazier2018}.

\subsection{Model parameterization}\label{sec_modelParameters}
Formulation of the GPR model and its domain (i.e., input) variables should be sufficiently complex to capture and describe the target physical phenomenon, but should otherwise be kept  as compact as possible, as it has been well established that the efficiency of GPR  tend to degrade in dimensions much higher than about 10 \cite{10.1145/3545611}.

In our context, dislocation densities 
representing the total lengths of dislocations per unit volume
of the twenty four $\langle$111$\rangle$\{110\} slip systems
are the obvious candidates for dimensional reduction for several reasons.
While coupling between densities of separate slip systems
is known to have a critical influence on slip system strain rates\cite{BULATOV20255611, akhondzadeh2021slip}, and thus should not be omitted entirely from a flow rule model domain,
a full specification of the densities on every slip system is not necessary to characterize these variations.
Conventional methods of reducing the number of slip system densities involve phenomenologically-derived coupling constants which are used to calculate lower dimensional representations of the complete density specification in the frame of a given slip system\cite{LEE1999625,AKHONDZADEH2020104152}.

With these issues in mind, we've chosen to  restrict the model's inputs to only two dislocation density parameters, being that of a particular $\langle$111$\rangle$\{110\} slip system $\rho_{i}$ and the total dislocation density over all slip systems $\rho_{\text{total}}$. This corresponds to the simplest form of flow-rule used in Crystal Plasticity models, encoding Taylor hardening
\cite{taylor1934mechanism}
Kocks-Mecking formalism
\cite{kocks2003physics}
for evolution of dislocation density.
 This choice preserves the
ability of the acquisition function to suggest configurations which may have strong coupling between slip system dislocation densities, in which $\rho_{\text{total}}$ is comparably greater than $\rho_{i}$, or configurations dominated by a single slip system where $\rho_{\text{total}}$ is approximately $\rho_{i}$, 
while reducing the total volume of the parameter space.
It should be noted, however, that reducing completeness of the specification by combining dislocation densities into a single parameter potentially induces noise as there will multiple ways to instantiate the same total dislocation density, each of which could yield a different strain-rate.

The remaining simulation parameters thought to determine strain rate are those describing the mechanical load which are summarized by
the six unique applied stress tensor components, $\sigma_{11}$, $\sigma_{12}$, $\sigma_{13}$, $\sigma_{22}$, $\sigma_{23}$, $\sigma_{33}$, defined with respect to a local coordinate system aligned with the slip plane and the slip direction to comply with equivalence among $\langle$111$\rangle$\{110\} slip systems.
Retaining all six of these dimensions as free parameters, which may take on either positive or negative values, 
allows the model to capture the non-Schmid effects observed in experiments\cite{christian1970dislocations,christian1983some} and predicted in molecular dynamics simulations of BCC materials
\cite{BERTIN2023119336,Kraych2019,Dezerald_2016}.
Our final GPR model is hence formulated using an eight-dimensional slip-system specific domain parameterization:
$(\boldsymbol{\sigma}_{i},\rho_{i},\rho_{\text{total}})$ where the density $\rho_{i}$ is specific to a particular slip system, and stress tensor $\boldsymbol{\sigma}_{i}$ is a rotation of the global stress $\boldsymbol{\sigma}$ into a coordinate system corresponding to that particular slip system in which the slip system's normalized Burgers vector and slip plane normal vector serve as the first and third vectors of an orthonormal basis, respectively.
The slip systems are enumerated in Table~\ref{table_ss_enum} and illustrated in Figure~\ref{rhombic_dodecahedron}.

\subsection{Details of the GP implementation}\label{sec_GPdetails}
The GP is implemented using the SingleTaskGP model from BoTorch using a Mat\`{e}rn covariance kernel as implemented by GPyTorch with homoscedastic noise and fixed smoothness parameter $\nu$=2.5, along with a Gaussian likelihood with gamma prior. 
Fitting of the model to training data is performed after first initializing length-scale hyper-parameters to values 
representative of the anticipated sensitivities of the objective to changes in corresponding model domain parameters.
For example, strain rate is anticipated to be highly sensitive to resolved shear stress as represented by the stress component $\sigma_{13}$ in the model's rotated frame and to the dislocation density of the given slip system, and thus ought to have shorter length scale hyper-parameters along those dimensions.
During model fitting, these initial values are optimized along with a scalar noise hyper-parameter which embodies a prediction of variance or noise in the outcome and is reported as a predictive likelihood variance or standard deviation in what follows.
Allowing hyper-parameters to be optimized during the fitting procedure in this way is sometimes referred to as the empirical Bayes approach \cite{dunlop2020hyperparameter,doi:10.1137/19M1284816}.
Having chosen to use homoscedastic noise, the noise hyper-parameter, length-scale hyper-parameters, and likelihood variance will be scalars which do not vary within the input domain.

\subsection{Acquisition function}\label{sec2Acqf}
Determining 
optimal microstructural and loading parameters used to instantiate 
simulations 
is the task of 
a custom-designed acquisition function.
This acquisition function is designed to 
sequentially identify simulation 
parameters where explicit simulation results are anticipated to maximally reduce the 
model uncertainty, which is equated with the GP model's 
posterior variance 
as described in Section~\ref{sec_modelSelection}.
A second requirement is the ability to identify multiple simulations (i.e., a so-called batch in the BO parlance) that can be executed concurrently while avoiding excessive inter-task redundancy. This batched mode of operation enables
reduction in the overall time needed to obtain the required simulation data through the use of parallel computing resources.

Unlike conventional Bayesian optimization in which the goal is to identify the maximum value of an expensive function, our goal is to optimize the fitting of a Gaussian process model over a prescribed domain by reducing its uncertainty.
Conventional acquisition functions like the well known expected improvement (EI) or knowledge gradient (KG) are thus not easily adaptable.
Instead, to reduce the model's total uncertainty we seek new training data at points where the GP model has highest posterior variance. Updating the model with new information at these points has been argued to be equivalent to reducing the Shannon entropy of the random variable which the model represents and has been previously applied in other works of optimal design of computer experiments \cite{currin1991bayesian,shewry1987maximum}.

A non-trivial challenge 
that arises
in our context is to
design an acquisition function which accommodates the difference in dimensionality between the eight dimensional GP model domain and the thirty dimensional DDD simulation domain consisting of microstructural
and loading parameters.
Due to the equivalence of the twenty four $\langle111\rangle\{110\}$ slip systems in BCC crystals, each point in the DDD simulation domain uniquely determines a collection of twenty four points in the GP model domain, one for each slip system.
This equivalence places constraints between the members of each subset of model domain points that correspond to a single simulation domain point,
specifically 
that they must all have stress tensors $\boldsymbol{\sigma}_{i}$ which are equivalent through a transformation from the global simulation frame to their individual slip system frame: $\boldsymbol{\sigma}_{i}=\mathbf{R}_{i} \boldsymbol{\sigma} \mathbf{R}_{i}^{-1}, \forall i \in [1,...,24]$, where $\mathbf{R}_{i}$ 
is the rotation matrix from a global Cartesian coordinate system to the coordinate system having basis $\{\hat{b}_i,\hat{n}_{i}\times\hat{b}_{i},\hat{n}_{i}\}$. 
Secondly, the sum of all slip system densities within the simulation domain must be equal to the total density parameter
of the model domain:
$\rho_{\text{total}}=\sum_{i=1}^{24}\rho_{i}$.

In addition to the above constraints, required by the  dimensionality mismatch between model and simulation, 
a mechanism to prevent degeneracy between model domain points is also desired.
For this, we've implemented a method of sequentially identifying points in the model domain 
under constraints which evolve with each proposed candidate.
This is achieved by reconditioning the model on every 8-D candidate as soon as they are proposed,
using the model's posterior mean as if it were an accurate prediction.
This reconditioning temporarily reduces the model 
posterior
variance in the neighborhood of those proposed points, thus reducing the propensity for the acquisition function to repeatedly propose similar points within a single batch~\cite{NEURIPS2018_498f2c21,lee2020costaware}.  
By preserving the fantasized model throughout the loop which identifies a single batch, similarity is dissuaded between all of the corresponding 8-D model domain points of {\em{the entire batch}}, considering all slip systems within the same configuration as well as slip systems present in different configurations that are part of the same batch 
to produce a batch of experiments that are globally maximally informative and non-redundant.
The flow of the acquisition function is outlined in Algorithm~\ref{alg_acqf}.

Due to the dimensionality and potential roughness of the search landscape, a multi-start optimization scheme from the BoTorch software package \cite{DBLP:journals/corr/abs-1910-06403} is used in which 
collections of random starting points are produced via Monte Carlo Sampling (MCS).
Local maxima of model posterior variance near each starting point are then found via the sequential least squares programming (SLSQP) method, as implemented in the SciPy software package \cite{2020SciPy-NMeth}.
After repeating the multistart optimization a specified number of times
and selecting the point for which the model posterior has maximum variance within the results, 
the point representing the first slip system in the model domain $(\boldsymbol{\sigma}_{1},\rho_{1},\rho_{\text{total}})$
is obtained.
The fantasized model is then reconditioned on this point using the value of its own posterior mean.
During this routine of proposing and optimizing candidates in the model domain, 
constraints discussed in the next section are enforced to ensure that the candidates are consistent with  physical and practical limitations.
All but the final of the remaining slip system density specifications are then determined sequentially using the same multi-start optimization method but under repeatedly updated constraints and reconditioning of the fantasized model.
These constraints restrict
model domain candidates to stress tensors which are transformed by the aforementioned rotations and values of density that must lie between a minimum density $\rho_{\text{min}}$
and an appropriate fraction of the remaining available density, $\rho_{\text{avail}}/(24-i)$ for the $i^{\text{th}}$ slip system where $\rho_{\text{avail}} = \rho_{\text{total}} - \sum_{j=1}^{i-1} \rho_{j}$, to allow for each remaining slip system to be assigned a density of at least $\rho_{\text{min}}$.
The final slip system is eventually assigned all of the remaining available density, resulting in a collection of densities $\{\rho_{i}\}$ whose total $\rho_{\text{total}}$ was determined by the first step.
In this way all thirty parameters in the domain of a single simulation target are chosen: $(\boldsymbol{\sigma},\vec{\rho})$. The temporary reconditioned fantasy model is preserved until all $N_{\text{batch}}$ targets in a batch of simulations are specified, to encourage variety within each batch.

\begin{algorithm}
   \caption{Acquisition function}\label{alg_acqf}
   \begin{algorithmic}[1]
      \State Create a copy of the GP model for fantasizing
      \For{ simulation $n=1$ to $N_{\text{batch}}$}
         \State Use MCS and SLSQP to find $(\boldsymbol{\sigma}_{1}, \rho_{1}, \rho_{\text{total}})$ subject to the constraint: $\rho_{\text{total}} - \rho_{1} \ge 23 \rho_{\text{min}}$
         \For{ slip-system $i=2$ to $23$}
            \State Evaluate $\rho_{\text{avail}} = \rho_{\text{total}}-\sum_{j=1}^{i-1}\rho_{j}$
            \If{$\rho_{\text{avail}} > (24-i)\rho_{\text{min}}$}
               \State Evaluate the model posterior mean $\dot{\gamma}_{i-1}$ at $(\boldsymbol{\sigma}_{i-1}, \rho_{i-1}, \rho_{\text{total}})$
               \State Recondition the fantasized GP model as if $\dot{\gamma}_{i-1}(\boldsymbol{\sigma}_{i-1}, \rho_{i-1}, \rho_{\text{total}})$ is an accurate prediction
               \State Use MCS and SLSQP to find $(\boldsymbol{\sigma}_i, \rho_{i})$ constrained by: $\rho_{\text{min}} < \rho_{i} < \rho_{\text{avail}}/(24-i)$ and $\boldsymbol{\sigma}_{i} = \mathbf{R}_{i}\mathbf{R}_{1}^{-1} \boldsymbol{\sigma}_{1} \mathbf{R}_{1} \mathbf{R}_{i}^{-1}$
            \Else
               \State $\rho_{i} \gets \rho_{\text{min}}$
            \EndIf
         \EndFor
         \State $\rho_{24} \gets \rho_{\text{total}} -\sum_{i=1}^{23}\rho_{i}$
         \State Append $(\mathbf{R}_{1}^{-1}\boldsymbol{\sigma}_{1}\mathbf{R}_{1}, \rho_{1}, \rho_{2}, ..., \rho_{24})$ to the batch of simulation parameters to be executed
      \EndFor
   \end{algorithmic}
\end{algorithm}
\subsection{Additional constraints}\label{sec2Con}
In preceding sections we described constraints 
upon simulation and model domain values
arising as consequences of their dimensional mismatch as well as part of the strategy of the acquisition function.
In this section we describe constraints designed to ensure physical and practical plausibility.

As the acquisition function assembles and scores candidate simulations,
candidates which don't meet  practical constraints are discarded.
While the acquisition function limits its parameter search to values of density below the remaining available density,
a minimum density per slip system is also enforced.
This minimum of 3.41$\times$10$^{11}$m$^{-2}$ was calculated to represent that of a single segment of a prismatic loop within the given box volume, ensuring that each slip system has an opportunity to produce non-trivial strain rate measurements.
The total density is then constrained above by 10$^{14.2}$m$^{-2}$ and below by the sum of densities of the first slip system and a minimum of 3.41$\times$10$^{11}$m$^{-2}$ per remaining twenty three slip systems.

The candidate stress component magnitudes are bound from above by $10^{8.7}$Pa
and their resulting global stress tensor is 
subject to a minimum eigen value (or eigen stress) 
constraint which ensures that the maximum magnitude of the eigenvalues of any applied stress tensor is at least $10^{7.5}$Pa.
This constraint was chosen to ensure that the load applied is  sufficient to produce usable measurements of plastic distortion and strain rate within a reasonable amount of compute time. 
While the constraints on applied stresses and dislocation densities limit the acquired data to relatively high driving forces they do not result in any loss of generality in the computational workflow and can be easily relaxed with more computational resources to target regimes with low driving forces. 
A more concrete exploration of distributions over the domain imparted 
by the constraints of the acquisition function is presented in Appendix~\ref{sec_intrinsic_bias}.
\subsection{Discrete Dislocation Dynamics}\label{sec2DDD}
Discrete dislocation dynamics simulations were performed using MoDELib \cite{po2014recent} with a parameterization for $\alpha$-Fe at 320~K similar to ref.~\cite{MAHLER2021100814}.
Simulation parameters influential to the characteristics of our results include the use of periodic boundaries and a system box size of approximately 1.5 $\mu$m per side, adaptive time steps with subcycling, and omission of cross-slip.
The mobility law employed which determines dislocation segment velocity incorporates non-Schmid effects and imparts a sensitivity to the $\sigma_{12}$ component of the applied stress tensor when specified in a slip system aligned basis \cite{PO2016123}.
As illustrated in Fig.~\ref{rhombic_dodecahedron}, the crystal has been rotated within the system box about the $[\bar{1}11]$ direction by 30 degrees to reduce the likelihood of annihilation of dislocations due to the periodic boundaries. 

Because the acquisition function specifies its desired target microstructures in terms of densities per slip system, we are left to choose the dislocation types, sizes, and locations.
To ensure every slip system may become active in a simulation, four randomly placed hexagonal prismatic dislocation loops of radius 1~$\mu$m were inserted into all simulations so that each slip system contains at least one segment of a prismatic loop.
The lower bounding constraints on density used by the acquisition function were adjusted accordingly to represent the existence of a single segment of a prismatic loop on each slip system.
To instantiate any remaining requested density per slip system, shear loops of radius 1~$\mu$m were inserted with centers at uniformly distributed random locations until that slip system's target density specification was satisfied.

After instantiating a targeted microstructure, batches of discrete dislocation dynamics simulations were performed under constant applied stress. To provide an interface between the DDD software written in C++ and the Python Bayesian optimization routines, a Python interface to MoDELib was implemented using pybind11~\cite{pybind11}.
The python interface provides functionality to specify and instantiate various microstructures, as well as to run DDD while measuring slip system resolved quantities of interest.
Specific to our purpose was the need to measure the
plastic strain per slip system $\gamma_{i}$, stress tensor $\boldsymbol{\sigma}$, and all slip system dislocation densities $\vec{\rho}$.

\subsection{Processing of measurements}\label{postprocessing}

Having obtained time-series measurements of the evolution of plastic strain and dislocation densities per slip system under a constant stress, $\gamma_{i}(\boldsymbol{\sigma},\vec{\rho}_{t})$, 
a scheme to properly filter and prepare the data for the Gaussian process model is needed.
First, the initial microstructures composed of perfectly circular or hexagonal loops  should first be relaxed under the desired loading condition. For this, a minimum threshold of 0.2\% strain accumulated on at least one slip system was imposed before data from any slip system was accepted for use in the model. In practice, however, low strain rate simulations
which would not satisfy this constraint within reasonable simulation times were highly sought after by the acquisition function and it was wasteful to entirely discard them when they failed to meet this strain threshold in a practical simulation time. 
To ensure that measurements from low strain rate simulations were represented in the data set, only the final state of any such simulations was accepted into the training data.

Second, reactions among dislocations and details of numerical implementation introduce statistical noise on the measured quantities.
To reduce the influence of these noise sources, we calculate the strain rate as the slope of a line fit to clusters of 100 consecutive measurements of plastic distortion, as counted from the end of the simulation. Corresponding values of slip system densities $\rho_{i}$, total density $\rho_{\text{total}}$, and time are obtained by averaging over the 100 consecutive measurements.
An additional consequence of these noise sources was the infrequent presence of outlying data in which changes to plastic distortion were  in the opposite direction of the applied load.
In the context of producing a flow rule for CP, these data points are considered unproductive deviations from the physics which it aims to capture.
To reduce the influence of this noise on the quality of the model, the strain rates of any measurements having $\tau_{rss} * \dot{\gamma} < 0$ were omitted from the training data set.

Providing too many similar data points to the model can induce negative effects like over-fitting.
Due to our choice to apply a constant mechanical load, the stress tensor values occupying six of our eight model domain dimensions are constant during a single simulation. 
Thus, accepting a large number of measurements from any single simulation may produce a data set containing dense clusters which are unevenly distributed over the model domain. Such redundant data points provide limited novel information while increasing the cost of training and querying the GPR model.
To counteract this issue, we restrict the total number of data points acquired per simulation to 240 by randomly selecting 10 from each slip system.

Exponential trends in the measured data (e.g., dislocation densities ranging from $10^{12}~m^{-2}$ to $10^{14}~m^{-2}$) can produce poorly fitted models when fed directly into conventional GP models. 
Additionally, non-uniform distributions in both model domain and model objective values will influence the magnitude of uncertainty of the model, producing higher uncertainty in more sparsely distributed regions, thereby increasing the frequency with which corresponding regions of the model domain are requested by the acquisition function.
A preconditioning transformation from measurement values into model data is thus preferred so that the GP model receives more uniformly distributed data in both the model domain and model objective.

Dislocation density is transformed by applying a $\log_{10}$ function to the total and particular slip system densities while assigning 0 to any value less than or equal to $1~ m^{-2}$ before presenting them to the model.
The six unique stress tensor components are similarly transformed to a $\log_{10}$ scale with the interval $[-1,1]$ Pa mapped to 0, but in a way which preserves the sign of the value, specifically 
\begin{equation}
\sigma^{\text{model}}_{ij} =
\begin{cases}\text{sign}(\sigma^{\text{meas.}}_{ij}) \log_{10}|\sigma^{\text{meas.}}_{ij}|,&|\sigma^{\text{meas.}}_{ij}|>1 \\
0,&|\sigma^{\text{meas.}}_{ij}|\leq1
\end{cases}
\end{equation} 
During acquisition function optimization, we impose both linear and non-linear constraints upon and among the stress and density target values as described in Section~\ref{sec2Acqf} to avoid requests for unobtainable simulation conditions.

While  the distribution of observed strain rates can also span  several orders of magnitude (e.g., from $0~s^{-1}$ to $10^{7}~s^{-1}$),
a purely logarithmic treatment of measured strain rate data would have imparted an inappropriately large representation of low strain rates within the entire range of measurements.
This was addressed by transforming low magnitude strain rate measurements linearly and high magnitude strain rates logarithmically via the piece-wise continuous $C^1$ function shown in Equation~\ref{EqStrainRateLog} that is applied to the magnitude of the measured strain rates. In this equation, $\dot{\gamma}_{\text{knee}}$ defines the boundary between linear and logarithmic behavior.
In the present work, we made the uninformed decision to use a value of $\dot{\gamma}_{\text{knee}} = 10,000~s^{-1}$, so that strain rates whose magnitudes were above this value were compressed logarithmically.
The sign of each measured strain rate was preserved and reapplied after this transformation.

\begin{equation}\label{EqStrainRateLog}
f(|\dot{\gamma}|) =
\begin{cases}
|\dot{\gamma}| & |\dot{\gamma}| < \dot{\gamma}_{\text{knee}} \\
\dot{\gamma}_{\text{knee}}(\ln|\dot{\gamma}|-\ln\dot{\gamma}_{\text{knee}}+1) & |\dot{\gamma}| \ge \dot{\gamma}_{\text{knee}}
\end{cases}
\end{equation}

After transforming the measured data into more uniformly distributed scales, the values in each dimension were rescaled in the unit interval [0,1] in each domain dimension, while the strain rate values are transformed into a standardized normal distribution $\mathcal{N}$(0,1).
\subsection{Workflow and automation}\label{sec2workflow}
\begin{figure*}[t]
   \centering
   \begin{center}
    \includegraphics[width=1.0 \textwidth]{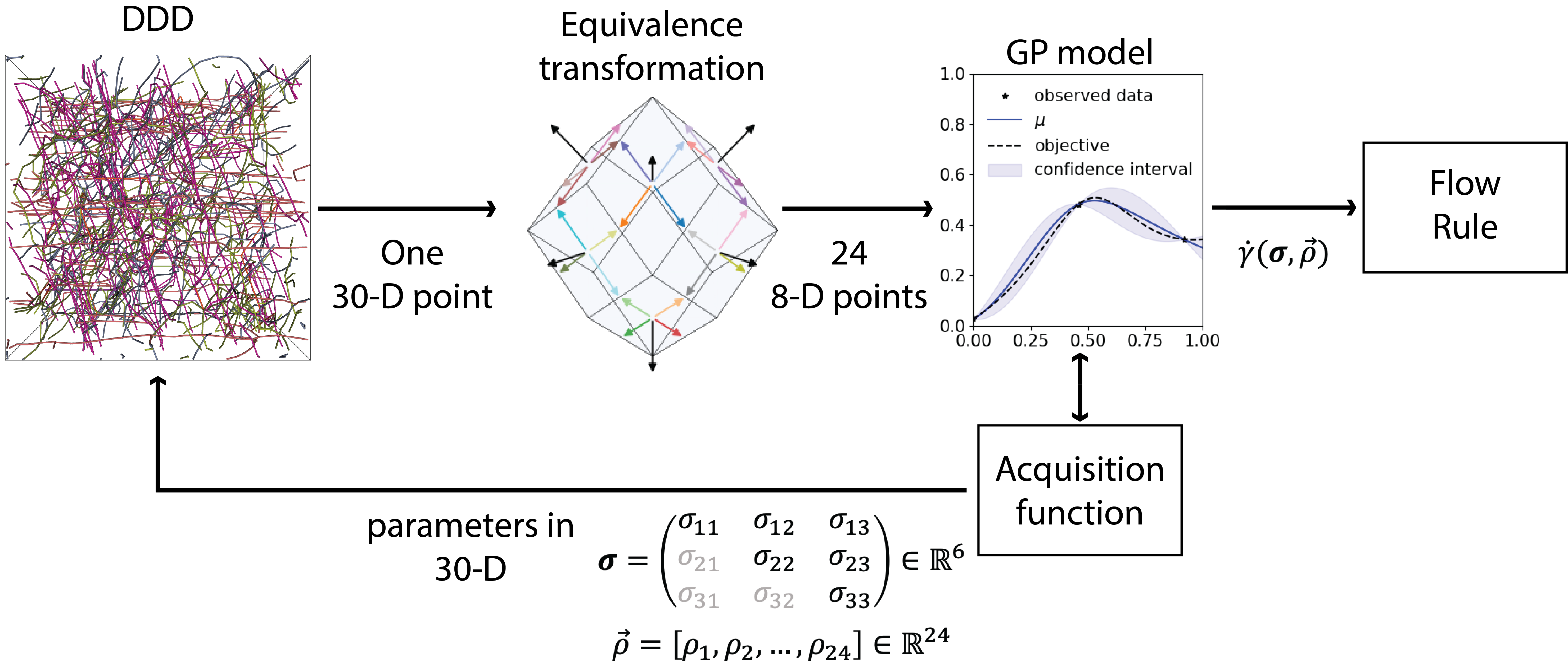}
   \end{center}
   \caption{Flow diagram illustrating our application of Bayesian optimization to guide parameter space exploration 
   and data generation
   from discrete dislocation dynamics (DDD)
   for use in a crystal plasticity flow rule $\dot{\gamma}(\boldsymbol{\sigma},\vec{\rho})$. Here $\dot{\gamma}$ is the strain rate of a slip system, $\boldsymbol{\sigma}$ is applied stress tensor, and $\vec{\rho}$ is the collection of dislocation densities per slip system.}\label{fig_flowdiagram}
\end{figure*}

The overall flow of data described by the preceding subsections
is illustrated in
Fig.~\ref{fig_flowdiagram}.
The 1-D diagram labeled as the GP model is a simple representation of Gaussian process regression and should not be mistaken for the actual model which is constructed over an 8-D domain.
Management and automation of the outermost active learning loop was implemented using Python, within which Parsl \cite{10.1145/3307681.3325400} was used to automate interactions with the SLURM workload manager in a way which minimized human interaction.
To ensure the sanity and productivity of each batch of simulations, stopping points were inserted immediately prior to instantiating a batch of simulations, as well as after each batch of simulations concluded.
Because a SLURM allocation was only allotted a restricted number of hours, Python and Parsl were also used to automatically restart each simulation after an allocation expired for a total of 5 allocations of 16 hours per simulation in a batch of 10 simulations per iteration of the outermost loop.
\section{Results and Discussion}\label{sec3}
\subsection{Discrete dislocation dynamics simulations}\label{sec3DDD}
\begin{figure*}[t]
   \centering
   \begin{center}
      \includegraphics[width=0.34 \textwidth]{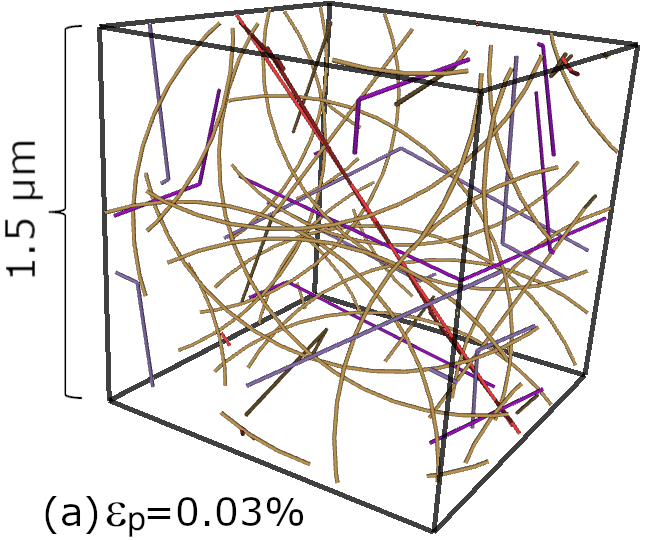}
      \includegraphics[width=0.34\textwidth]{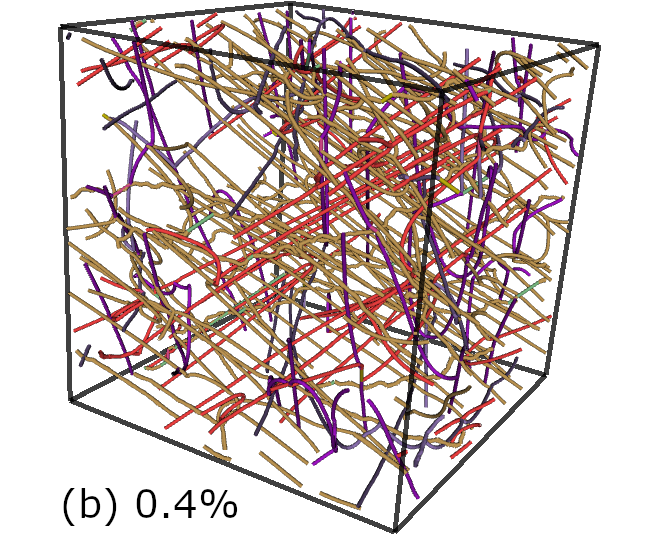}
      \includegraphics[width=0.34 \textwidth]{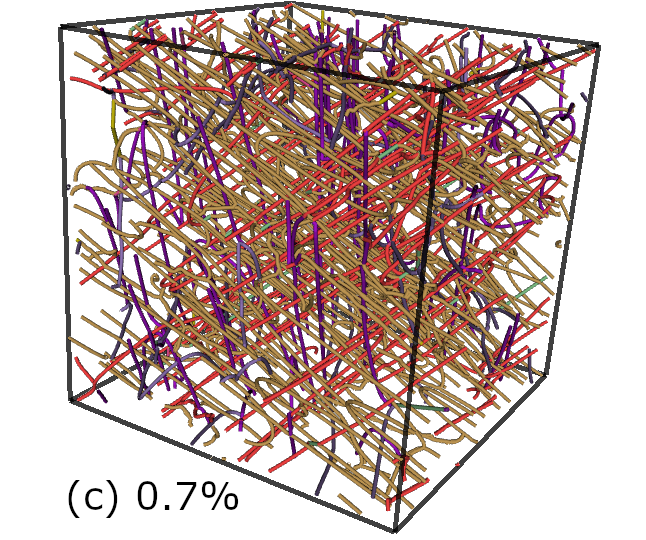}
      \includegraphics[width=0.34\textwidth]{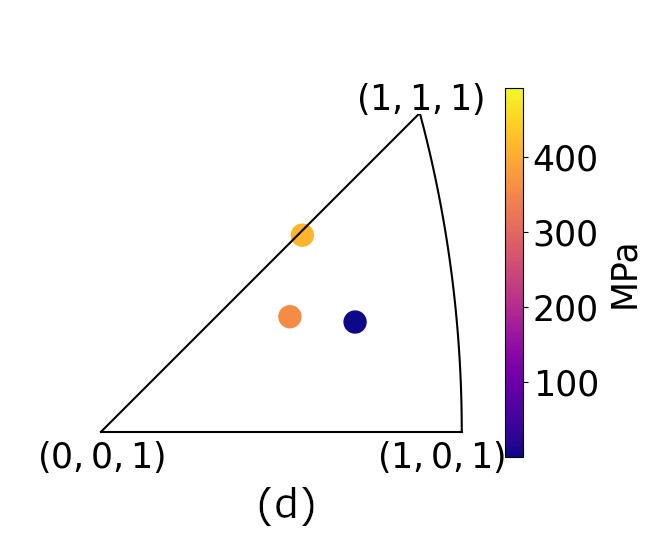}
   \end{center}
   \caption{Evolution of a dislocation dynamics simulation 
   (a-c) under the load of a constant stress tensor whose eigen vectors and eigen values are shown relative to crystal orientations in the projective triangle (d). Measurements of equivalent plastic strain percentage are annotated as $\varepsilon_{p}$ while slip system resolved quantities corresponding to images (a-c) are shown in Fig.~\ref{fig_example_simulations}a.
   }\label{fig_ddqt_6d512bca}
\end{figure*}

\begin{figure*}[t]
   \centering
   \begin{center}
      \includegraphics[width=\textwidth]{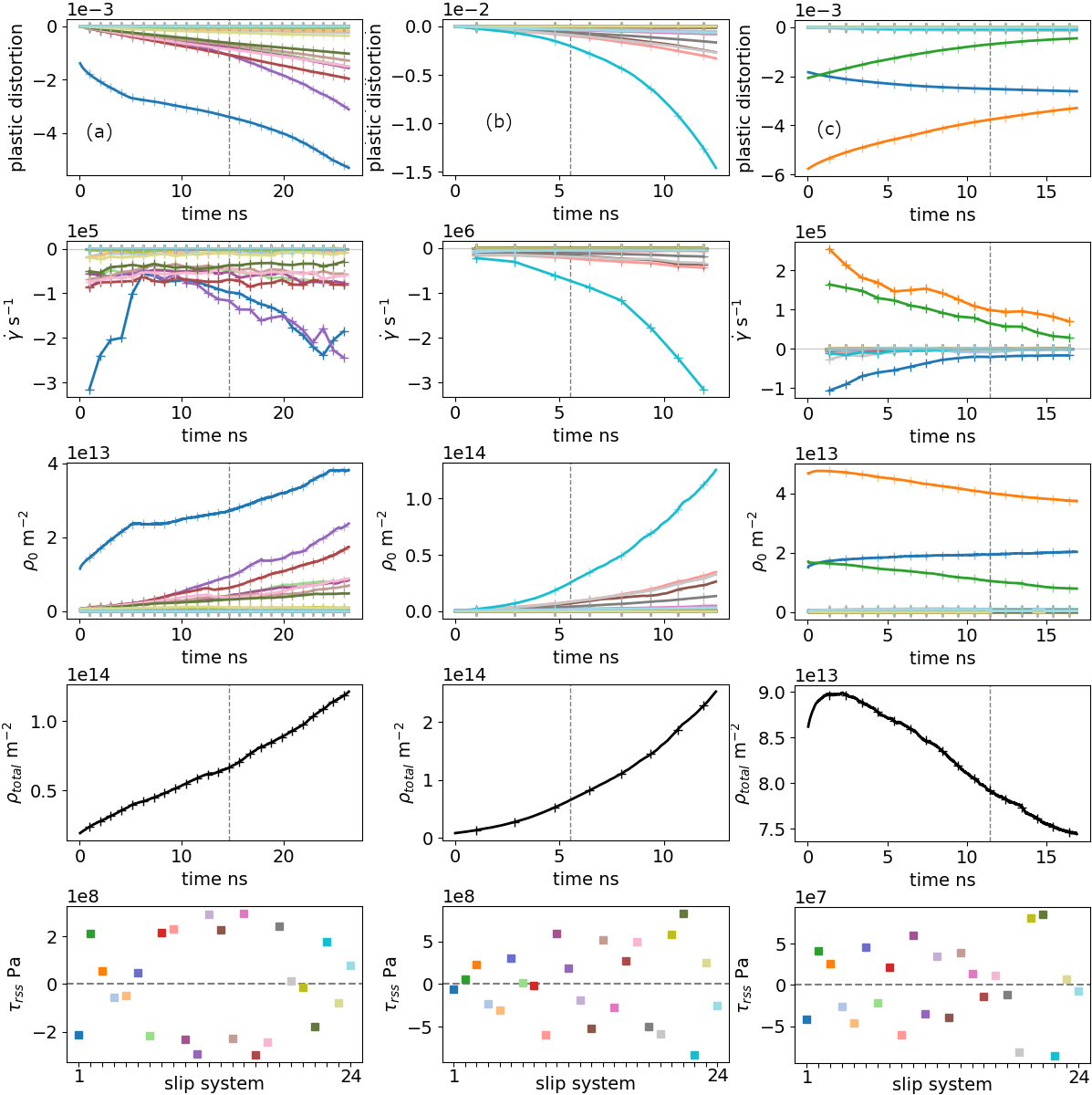}
   \end{center}
   \caption{A variety of the resulting simulation measurements
   of plastic deformation, strain rate ($\dot{\gamma}$), particular slip system densities ($\rho_{0}$), total dislocation density ($\rho_{total}$), and resolved shear stress ($\tau_{rss}$).
   Each column corresponds to an individual DDD simulation instantiated by the acquisition function. The colors 
   consistently
   correspond to
   specific slip systems 
   as enumerated in Table~\ref{table_ss_enum}
   while vertical dotted lines indicate the times at which the minimum strain condition is satisfied. 
   Thus it is evident that a variety of slip systems are activated in the simulations.
   }\label{fig_example_simulations}
\end{figure*}

Evolution of a typical DDD simulation is illustrated in Fig.~\ref{fig_ddqt_6d512bca} at three stages: (a) instantiation of 1 $\mu$m radius shear loops and prismatic loops as described in section~\ref{sec2DDD}, (b) evolution  
until a slip system 
reaches
0.2\% strain 
difference from its initial state,
and (c) final higher density configuration.
These images correspond to the simulation 
measurements 
displayed in
Fig.~\ref{fig_example_simulations}a.

The prominence of rising densities is a consequence of our use of periodic boundaries.
In a DDD convention in which plastic distortion
of a shear loop resolved on its own slip system
is always negative, applying a positive resolved shear stress $\tau_{\text{rss}}$ 
causes loops within that slip system to shrink, contributing 
positively to the strain rate, while a negative $\tau_{\text{rss}}$ will cause 
dislocation loops 
to grow and contribute negatively to the slip system's strain rate.
Thus the dislocation loops which contribute positively to strain rates are short-lived, shrinking until their inevitable self annihilation and often do not persist far beyond the 0.2\% strain threshold. For this reason 
data from slip systems having primarily positive strain rates are not anticipated to be of imperative value to CP simulations, and 
the data sets produced by our simulations are
acceptably 
sparse in the regions of positive strain rates. Fig.~\ref{fig_example_simulations}c demonstrates an occasional example counter to this trend, in which the applied load and configuration led to a slow shrinking of the majority of dislocation loops,
specifically those on slip systems numbered 2 and 3, while the dislocation density on slip system 1 appears to have been stabilized under the load.

Fig.~\ref{fig_example_simulations} illustrates several other notable features found within our measurements.
Deformation within the simulation presented in the second column is dominated by
the
single slip system 
numbered 22
whose dislocation density and plastic deformation show consistent, if not accelerating, growth.
This is in contrast to the simulation of the first column in which the dislocation densities of 
slip systems
1 and 11, which have identical Burgers' vectors but differing glide planes,
are growing at rates which vary in time,
presumably
due to interactions between dislocations on 
the
differing 
slip systems.
It could be argued that the initially high strain rate of the first slip system was due to lack of interaction with the other slip systems resulting in a period of free movement which was soon to be suppressed when interactions became prominent, producing varying strain rates and noise in the functional relationship $\dot{\gamma}(\boldsymbol{\sigma},\vec{\rho})$.
The leading theory of how
slip system interactions reduce strain rates is via the formation of junctions between dislocations whose mobilities are reduced or negligible \cite{bulatov2006dislocation,PhysRevLett.121.085501}.
In this particular example the two slip systems which appear to be interacting cannot form junctions because they have identical Burgers' vectors.
It should be noted that the specific configurational information which induces these interactions and the resulting strain rates is lost when the configuration is represented as dislocation densities per slip system, but could be recovered from the raw logs of the DDD simulations.
The significant variations in strain rate measurements for a given stress tensor and list of densities are then represented by the fitted noise hyperparameter of the GP model.

The non-static nature of dislocation densities produced by the dynamics described above results in differences between the measured model domain values and those desired by the acquisition function.
Generally, the longer a DDD  runs, the farther from the targeted densities the system wanders.
The resulting differences between targets and measurements are shown concretely by comparing Figures~\ref{fig_intrinsic_distribution}(b,e) or (e,f) in the Appendix.
Additionally, the lack of precise control over the evolution of dislocation densities in DDD simulations prohibit deployment of a heteroscedastic noise model by impeding the required measurements of noise through resampling of identical domain values. Note however that the GP model is trained using the actual density and stress values measured during the DDD simulations, not to the target values. While the drift from the target values somewhat diminishes the accuracy of the BO at targeting specific regions of configuration space, it does not introduce a bias in the model. 

\subsection{Model evolution}\label{sec3model}
As iterations of the sequence of simulations proceed, producing new data to which the model is fitted and new simulations are instantiated, several pertinent questions arise.
When should the sequence of simulations stop? How accurate is the model?
Which of the model parameters is strain rate most sensitive to?
These questions are answered by examining the trends in model posterior uncertainty, comparing its predictions to a collection of independent test measurements, and by inspecting model hyper-parameters, respectively.

\subsubsection{Convergence on the model scale to test data}\label{sec_convergenceOnModelScale}
To generate an independent set of test data for comparison, we repurposed the results of 335 simulations obtained during the software development and testing phase of the project, 227 of which had stress tensor components distributed nearly uniformly on the normalized log scale of the GPR,
while 108 simulations followed a BO guided trajectory through parameter space using a different set of processing methodologies than those outlined in Section~\ref{postprocessing}.
From these simulations 47,097 data points were extracted by following the methods of Section~\ref{postprocessing}, with the exception that those simulations satisfying the minimum strain condition had all of their satisfying data accepted rather than being randomly down-selected.

\begin{figure*}[t]
   \centering
   \begin{center}
   \includegraphics[width=0.49\textwidth]{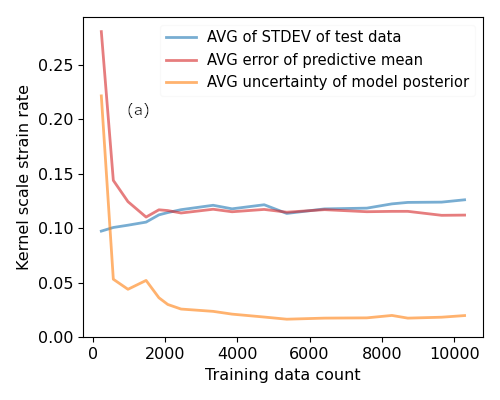}
      \includegraphics[width=0.49\textwidth]{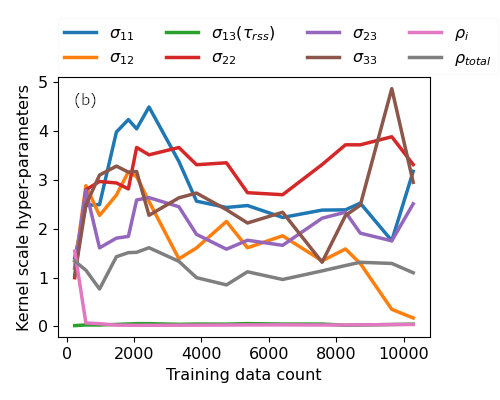}
      \end{center}
   \caption{Evolutions of (a) the model's posterior uncertainty (standard deviation) and predictive error 
   compared to the noise of the test data on the standardized scale of the kernel and (b) model hyper-parameters in units of distance on the scale used by the kernel.}\label{fig_convergence_plots}
\end{figure*}

It's anticipated that, for an accurate model, the error should converge to a magnitude similar to that of the intrinsic noise in the test data.
To evaluate noise in the test data,
a method of evaluating standard deviation over neighborhoods in the domain was needed, and thus some sort of spatial binning.
Unfortunately, the distributions of both test and training data sets in the model domain are highly skewed, clustered, and not uniform, leaving large volumes without representation.
We elaborate on these distributions in Appendix~\ref{sec_intrinsic_bias}.
Additionally, binning of the entire eight dimensional model domain would be computationally prohibitive, i.e. requiring a total of N$^8$ bins when splitting each dimension into N parts.
To address this we implemented an efficient algorithm of sparse binning which recursively splits the binned data, discarding empty bins as they arise until the desired bin size is reached.
Binning the test data in this way using a bin side length of $2^{-5}=0.03125$ on the scale of the normalized model unit hypercube, we then calculated the mean and standard deviation over each bin of strain rate in the units of the kernel scale, omitting bins having fewer than 8 points from the calculation of standard deviation.
With these calculated bin statistics of our test data we then probed the model at bin average domain values to calculate the corresponding
error of the model posterior predictive mean.
Evolution of the overall averages of these are shown in the leftmost plot of Fig.~\ref{fig_convergence_plots}.
If we denote the model posterior as $f$, then the curve labeled `AVG error of predictive mean' was calculated as
$
\langle|f(\langle\vec{x}_{\text{test}}\rangle_{\text{bin}}) -\langle\dot{\gamma}_{\text{test}}\rangle_{\text{bin}}|\rangle
$
where the outer average $\langle\cdot\rangle$ is over all of the bins, and the inner averages $\langle\cdot\rangle_{\text{bin}}$ are over domain $\vec{x}_{\text{test}}$ and strain rate $\dot{\gamma}_{\text{test}}$ test  data within individual bins.

Considering that our implementation of BO is designed with the goal of producing data
which reduces the overall total uncertainty in the model,
it follows that 
the desirability for new data 
may be equated with
the posterior standard deviation of the GP model.
Thus as the total uncertainty, or model posterior standard deviation, converges to zero, the worth of generating new training data also vanishes.
Note that this is not the same as the model's likelihood standard deviation, which is a prediction of how much the mean could vary due to noise,
and is a scalar in the case of homoscedastic models such as ours.

Figure~\ref{fig_convergence_plots}a confirms that the above trends have occurred on average.
The average uncertainty in the model approaches and remains near zero, while the average error found in the predictive mean of the model is in close proximity to the average noise (standard deviation) in the test data and begins dropping further between around 8,000-10,000 points.

The non-negligible magnitude of the noise in DDD strain-rate measurements induces a lower limit on the error of our model, forcing the average error of its predictions to be at least as bad as the noise.
This is shown by the similarity between the curves labeled `AVG of STDEV of test data' and `AVG error of predictive mean' in Figure~\ref{fig_convergence_plots}a.
This echos observations by another study\cite{AKHONDZADEH2020104152} which discussed the effect of noise on fitted coefficients of the Kocks-Mecking model using a large database of DDD simulations. 
We concur that methods of reducing and compensating for the noise obtained in these measurements are needed to improve model predictions.
While optimizing the reduced dimensional representation of all 24 slip system densities could possibly reduce noise, \emph{e.g.}~forest\cite{PhysRevLett.89.255508},  coplanar\cite{MA20043603, NGUYEN2021102940}, or glissile/sessile, our choice of a simpler specification avoided further complicating the acquisition function.

It might be noticed that noise in a constant set of test data should be constant, but varies on the scale presented here.
This occurs because the normalizing and standardizing transformations 
which optimize the data for the kernel are not constant, but evolve as the training data set grows.

\subsubsection{Evolution of the model's physical meaningfulness}\label{sec_convergenceTowardPhysicalMeaning}
It is also important to use prior knowledge of physical phenomena to assess the quality of the model, in particular we should validate that the model hyper-parameters correctly indicate the importance or irrelevance of particular features.
These sensitivities correlate with the model length scale hyper-parameters in a way such that 
smaller values correspond to model domain dimensions in which a small change will produce a large response in the value of predicted strain rate. 
The evolution of model length-scale hyper-parameters is shown in Fig.~\ref{fig_convergence_plots}b.

The small  hyper-parameters  observed for slip system resolved shear stress $\tau_{rss}=\sigma_{13}$ and dislocation density $\rho_{i}$ confirm that they are the features to which strain rate is most sensitive.
The total dislocation density is then consistently the next most influential feature, indicating a weak coupling between slip systems. In the final two iterations of the BO sequence the model discovers a greater sensitivity to the $\sigma_{12}$ stress tensor component, a result of non-Schmid effects. This is consistent with the specification of the mobility law which encodes the effect of a shear stress in the $\{ 110\}$ plane located at 60$^\circ$ to the slip plane in the $<111>$ zone.
By the final iteration the remaining unique stress tensor components are ranked nearly the same in regards to their lack of influence upon strain rate.

Due to the inverse relationship between feature sensitivity and hyperparameter values, fluctuations in the larger valued hyperparameters have lesser impact on model behavior due to their correspondence with features to which the model is least sensitive.
Although the larger valued hyperparameters do not tightly converge to final values, their fluctuations have the least impact, while the most sensitive features have hyperparameters which are consistently so small that they are difficult to distinguish when presented on the same scale.

Considering the aptness of length-scale hyper-parameter values in light of physical phenomena, the similarity between model predictive error and noise in the test data, and the low total uncertainty in the model posterior, it should be expected that generating additional training data would yield only minor improvements with regard to its quality on the scale of the kernel.

\subsubsection{Convergence on the physical scale}\label{sec_convergenceOnAPhysicalScale}
\begin{figure*}[t]
   \begin{center}
   \includegraphics[width=0.58\textwidth]{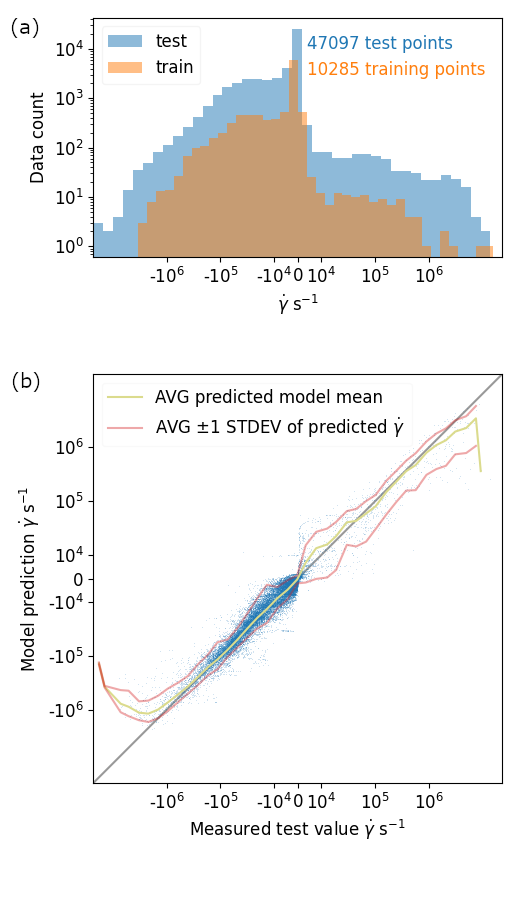}
   \end{center}
   \caption{Parity plot (b) and comparison between test and training sets (a) as of the 16$^{th}$ iteration. The averages of predictive mean and $\pm1$ predictive likelihood standard deviation about the mean are representative of the accuracy of and expected error of the Gaussian process model per strain rate. Histograms of strain rates in the training and testing data sets are shown in (a) to attribute regions of high STDEV and error to sparsity or absence of training data for corresponding strain rates.
   }\label{fig_parity_plot_with_histogram}
\end{figure*}

Assessing the quality of the model predictions on the scale of physical strain rates, rather than on the scale of the kernel discussed above, is best done using the parity plot shown in Fig.~\ref{fig_parity_plot_with_histogram}b.
This plot, presented with tick labels showing corresponding values on the physical scale, shows the strain rates predicted by the model as pixels above corresponding measured test values, which would ideally all fall upon the diagonal line.
Averages of these model predictions and model likelihood standard deviation, as binned on the intermediate log-like scale but averaged per bin on the physical strain rate scale, are shown as solid curves.
The average predicted model mean overlaps well with the ideal parity line over all of the training data, a histogram of which is shown above the parity plot in Fig.~\ref{fig_parity_plot_with_histogram}a.

The histograms of test and training strain rate values emphasize the difference in sizes of the two sets, as well as the rarity with which we obtained positive strain rates corresponding to shrinking dislocation loops, and the rarity with which we obtained very high magnitude strain rates.

Near regions where training data is sparse
it is appropriate that the average predicted model mean deviates from parity.
Similarly it is appropriate that the region enclosed by $\pm1$ standard deviation from the average of predicted $\dot{\gamma}$ widens above these sparsely populated training strain rate values.
As of the final iteration we calculate, on the scale of the kernel, that 63.9\% of all test measurement strain rates fall within $\pm1$ likelihood standard deviation from the corresponding predictive mean value, nearing the ideal value of 68\%.
An alternative perspective showing intervals of $\pm1$ likelihood standard deviation centered about the test measurement values is presented in Appendix~\ref{fig_parity_plot_with_histogram2}.

Overall, the final Gaussian process model produces predictions 
which, on average, agree well
with measurements at a physical scale.
Although, as the parity plot shows, it should be expected to have variations on the order of single digit factors.
Our extensive investigation of model quality has shown that there is little to gain from generating additional data. In contrast, it would be more beneficial to devise methods to reduce or compensate for the noise in DDD measurements 
as well as optimizing the domain transformations so that the model domain is more uniformly covered by the training data.

\section{Conclusion}\label{sec4}
In this work a self-guided computational pipeline has been demonstrated which instantiates simulations to generate prioritized training data for a machine learned model which bridges the scales of Discrete Dislocation Dynamics and Crystal Plasticity.
By exploiting the equivalence of BCC $<$111$>$\{110\} slip systems and reducing dimensionality of the model domain we've 
found a parameterization compatible with Crystal Plasticity 
which increases the yield of data from our simulations while avoiding poor model performance which would arise from the curse of dimensionality.
This required designing a custom acquisition function which iteratively fills information into under-specified dimensions with adaptively constrained target values chosen from the model domain to produce batches of higher dimensional targets describing the full loading and dislocation density parameters required to instantiate a batch of DDD simulations.
The model produced by training on data measured from a sequence of simulations instantiated by this realization of Bayesian optimization
converged in distribution to that of a notably larger set of test data within approximately 160 simulations,
reaching a state in which additional training data would provide little benefit to the accuracy of its predictions.
This demonstrates the efficiency and value of guiding simulations through the use of a model's uncertainty.
\backmatter
\bmhead{Acknowledgments}
Research presented in this paper was supported by the Laboratory Directed Research and Development program of Los Alamos National Laboratory under project number 20220063DR. This research used resources provided by the Los Alamos National Laboratory Institutional Computing Program, which is supported by the U.S. Department of Energy National Nuclear Security Administration under Contract No. 89233218CNA000001. G. Po acknowledges the support of LANL through the 2024 IMS Distinguished Faculty Scholar program.

\bibliography{bibliography}
\newpage
\bmhead{Supplementary information}
\begin{appendices}
\section{Distributions of simulation targets and eventual measurements}\label{sec_intrinsic_bias}

In Section~\ref{sec2Acqf}
we described how our acquisition function was designed to address 
the
differences in dimensionality between model domain and simulation parameters,
while in 
Sections~\ref{sec2Con}~and~\ref{postprocessing},
we described simulation constraints and transformations, respectively, intended to improve the distributions of data presented to the model in the spirit of regularization so that it may produce a better fit.
To assess the influence of these designs, transformations, and constraints we present here a study of the intrinsic distributions they collectively impart.
We compare these intrinsic distributions to the final collection of training data obtained from measurements to gain insight into the ability of our BO implementation to navigate parameter space under the influence of the intrinsic distributions.

\begin{figure*}[t]
   \centering
   \begin{center}
      \includegraphics[width=1.0\textwidth]{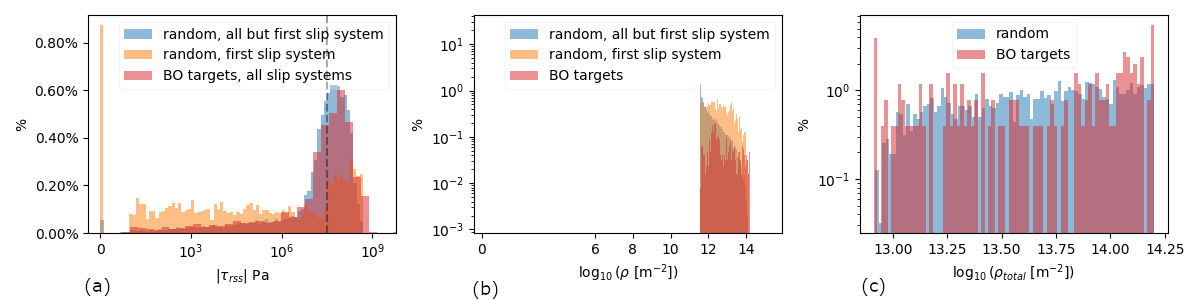}
      \includegraphics[width=\textwidth]{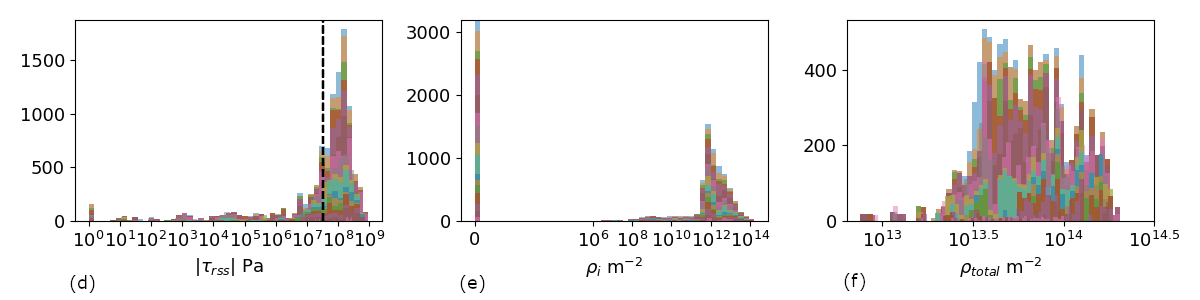}
   \end{center}   
      \caption{(a-c) Intrinsic distribution of model domain target values imparted by the constraints of the acquisition function compared to those obtained as targets during the BO sequence of simulations. Target specifications of the first slip system are separated to show the strong influence of the first phase of the acquisition function in which stress tensor and total density are not constrained by preceding slip system specifications.
      (d-f) Histograms showing the evolution and trends in measurements (not necessarily identical to the targets) of the most sensitive model domain values. The dashed line at $|\tau_{rss}|=10^{7.5}$~Pa indicates the constraint placed upon eigen stresses described in Section~\ref{sec2Con}. Colors correspond to iterations.}
\label{fig_intrinsic_distribution}
\end{figure*}

To characterize the intrinsic distributions of targeted model domain values
imparted by our constraints and methods of determining values for  under-specified dimensions,
we have removed the influence of the GP model upon the choices of the acquisition function, allowing the Monte Carlo sampler to sequentially determine 2,000 simulation domain targets ($\boldsymbol{\sigma}$, $\vec{\rho}$) through the application of Algorithm \ref{alg_acqf}.
The results of this process are prefixed with the label ``random" in Fig.~\ref{fig_intrinsic_distribution}(a-c).

Because the first phase of the acquisition function determines the stress tensor and total dislocation density to which all slip systems are constrained, it follows that there is a notable difference between the first slip system and all others
within the distributions of resolved shear stress and particular slip system densities.
The lower bound on maximum eigenstress per candidate
indicated by a vertical dashed line in Fig.~\ref{fig_intrinsic_distribution}a,d induced peaks in distribution just above it.
The first slip system, being less constrained, had a more uniform distribution over lower $|\tau_{rss}|$ and a very high peak at $|\tau_{rss}|=0$,
while the remaining slip systems, being constrained to fixed rotations of the already-determined stress tensor, had resolved shear stresses concentrated just above the eigenstress constraint value.
If the eigenstress constraint was absent, and arbitrarily slow simulations were allowed to be instantiated, we might guess that the intrinsic distribution of $|\tau_{rss}|$ for the first slip system would be uniform on this log scale, but $|\tau_{rss}|$ for the remaining slip systems would be a broader peak with a less severe skew toward higher values due to rotation of the stress tensor.

In the BO guided sequence of simulations the acquisition function used the uncertainty of the evolving GP model to guide its choices in the target specifications it produced, labeled here as BO targets.
The coarse similarity between the resolved shear stress of BO targets and the resolved shear stresses on all but the first slip system is due to the effective randomness of resolved shear stress on twenty three out of twenty four slip systems.
On a finer scale, there is a noticeable skew to the right in the $|\tau_{rss}|$ BO targets, consisting of a sharp peak and a low bulge protruding into higher $|\tau_{rss}|$.
A similar skew and sharp peak is present in the distribution of measured resolved shear stress (see Fig.~\ref{fig_intrinsic_distribution}d), but it lacks the broad character of the intrinsic distributions.
This might be attributed to differences in productivity between high and low stress, and thus high and low strain rate, simulations.

The distribution of BO targets for particular slip system densities $\rho_{i}$ resemble neither distribution of the first or the remaining slip systems particularly well, indicating that the 
acquisition function had greater liberty in guiding slip system density specifications.
However, the distribution of measured slip system density values which evolved during the simulations more closely resemble the distribution labeled ``random, all but first slip system,"
with the exception of the presence of a large number of slip systems having $\rho_{\text{min}}$ dislocation density.
So while the acquisition function had the most liberty to determine dislocation density targets, their evolution during the simulation caused measurements to drift significantly far from their targets.

Total dislocation density was the least constrained and manipulated target parameter, yielding the more uniform distribution labeled ``random" in Fig.~\ref{fig_intrinsic_distribution}c.
The targets produced by our BO sequence then had a fairly rough, almost uniform distribution.
The resulting measured values of total dislocation density, similar to the measurements of particular density, were highly modified by their evolution during simulation.
The lack of low total dislocation density measurements is perhaps due to the need to satisfy a minimum strain threshold, while the low count of particularly high total densities could be due to a corresponding slower deformation as well as higher computational load.

From the distributions of measurements, it's clear that there are large regions of the model domain which lack training data, specifically intermediate values above zero and below the practical thresholds of resolved shear stress or dislocation density.
Because the dynamics corresponding to values of zero in these dimensions are physically important, a simple restriction of the model domain to higher values is not preferable.
With these observations in hindsight, in the future it might be beneficial to design a transformation which contracts this nearly empty intermediate valued volume and distributes the measured values more evenly in the model domain.

\section{Alternative parity plot}
\begin{figure*}[t]
   \begin{center}
      \includegraphics[width=0.5\textwidth]{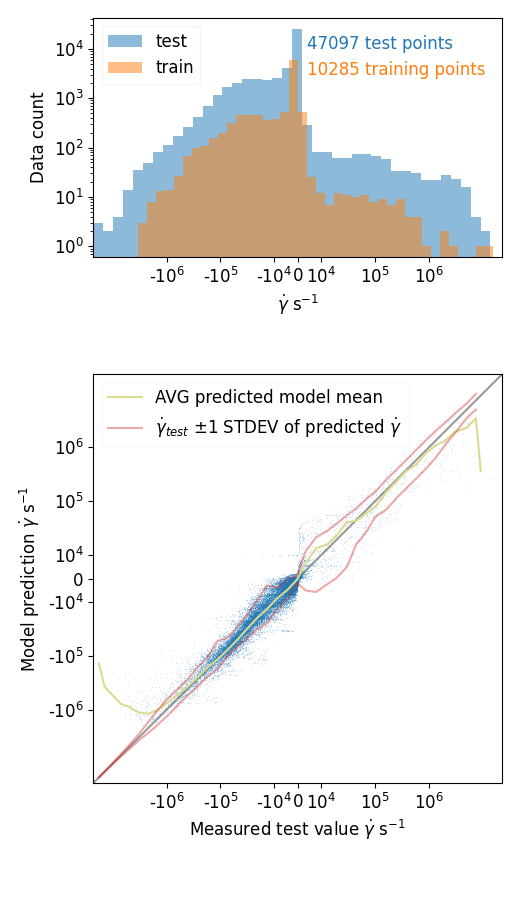}
   \end{center}
   \caption{Parity plot and comparison between test and training sets as of the 16$^{th}$ iteration. This figure differs from Fig.~\ref{fig_parity_plot_with_histogram} in that the interval of $\pm1$ STDEV of the likelihood is centered around the values of test strain rate, rather than the predictive mean of the model.}\label{fig_parity_plot_with_histogram2}

\end{figure*}

By centering the bounds of $\pm1$ STDEV of predicted $\dot{\gamma}$ about the test values $\dot{\gamma}_{test}$, Fig. \ref{fig_parity_plot_with_histogram2} shows how well the predicted model likelihood standard deviation follows the prediction error shown in the distance of each blue pixel from the axis.
\section{Sequential evolution of parity}
\begin{figure*}[t]
   \begin{center}
   \includegraphics[width=0.2\textwidth]{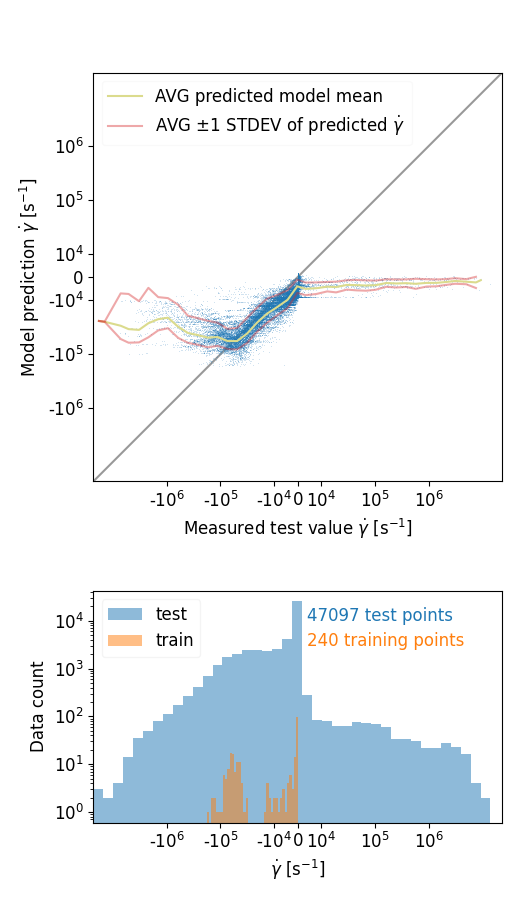}
\includegraphics[width=0.2\textwidth]{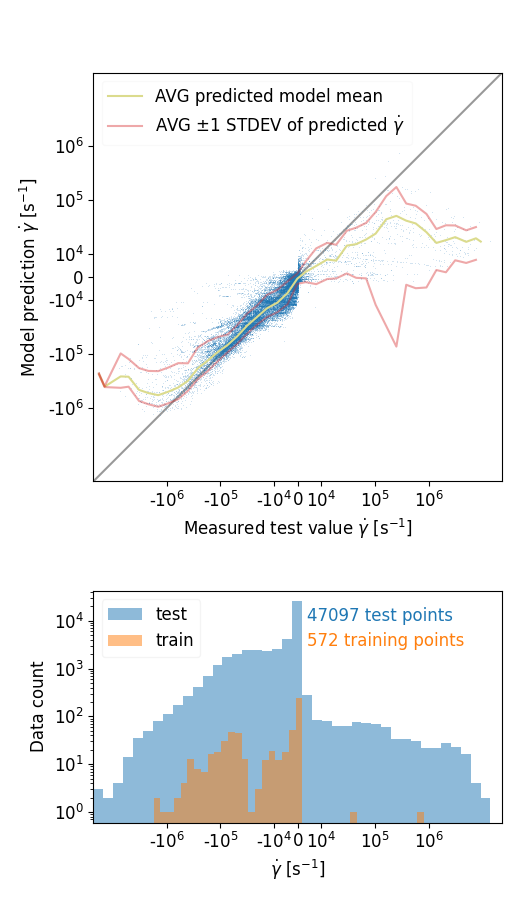}
\includegraphics[width=0.2\textwidth]{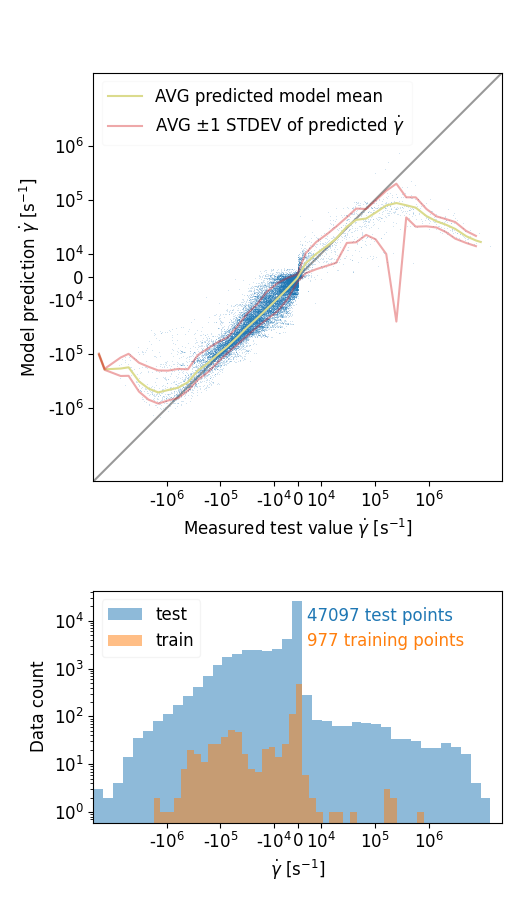}
\includegraphics[width=0.2\textwidth]{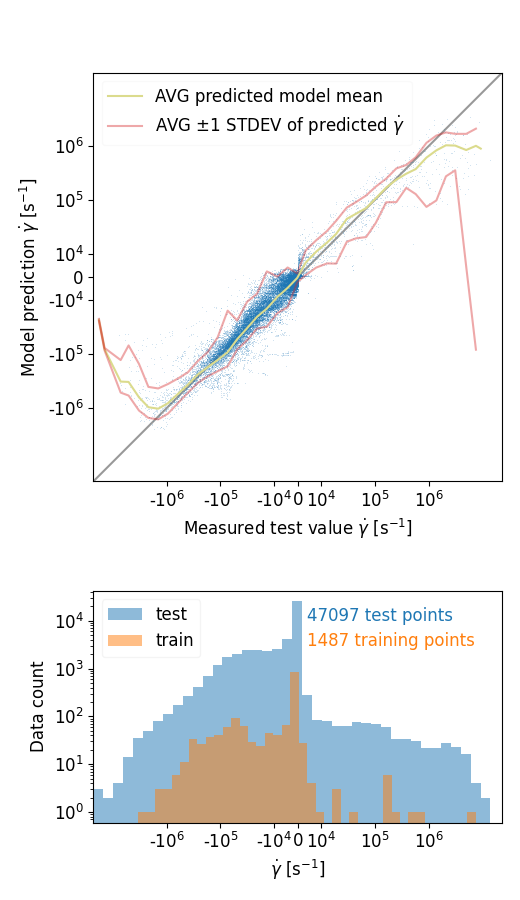}
\includegraphics[width=0.2\textwidth]{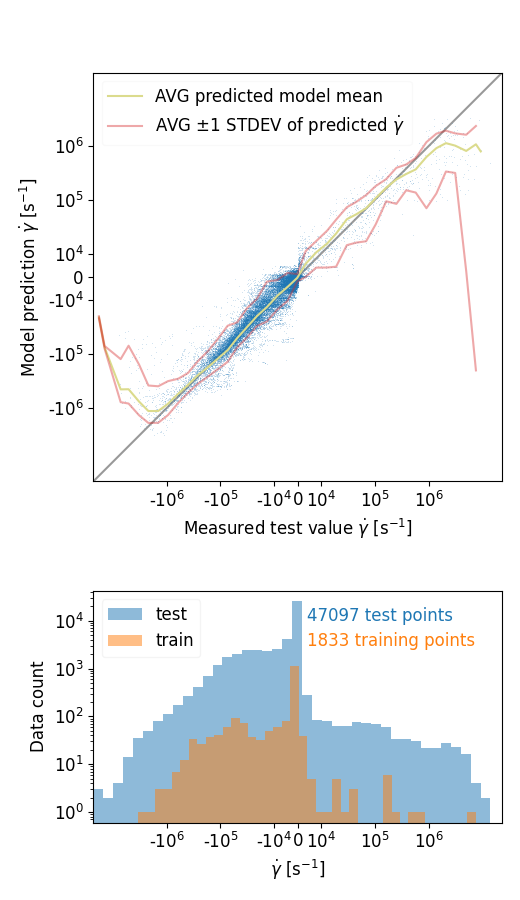}
\includegraphics[width=0.2\textwidth]{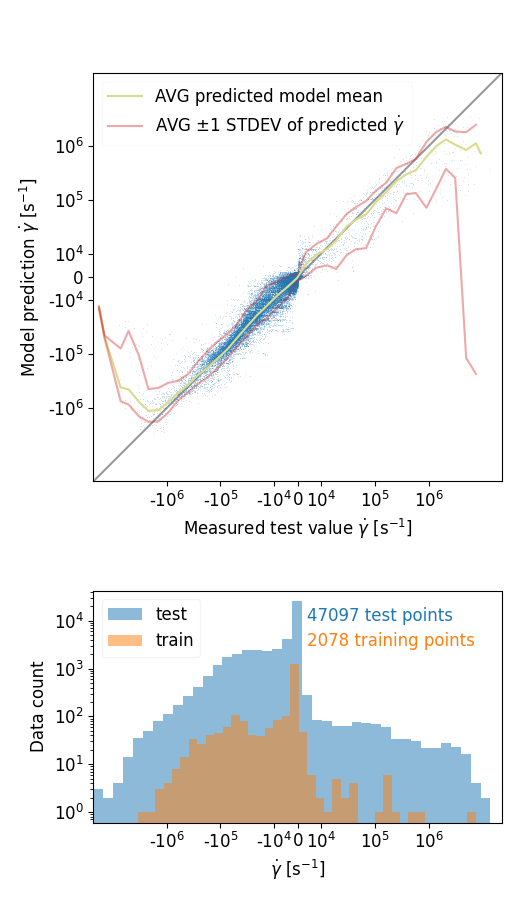}
\includegraphics[width=0.2\textwidth]{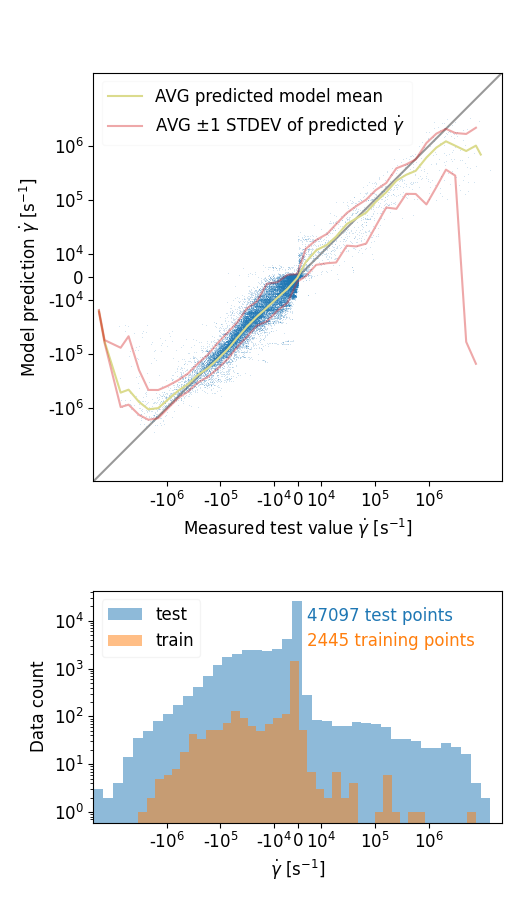}
\includegraphics[width=0.2\textwidth]{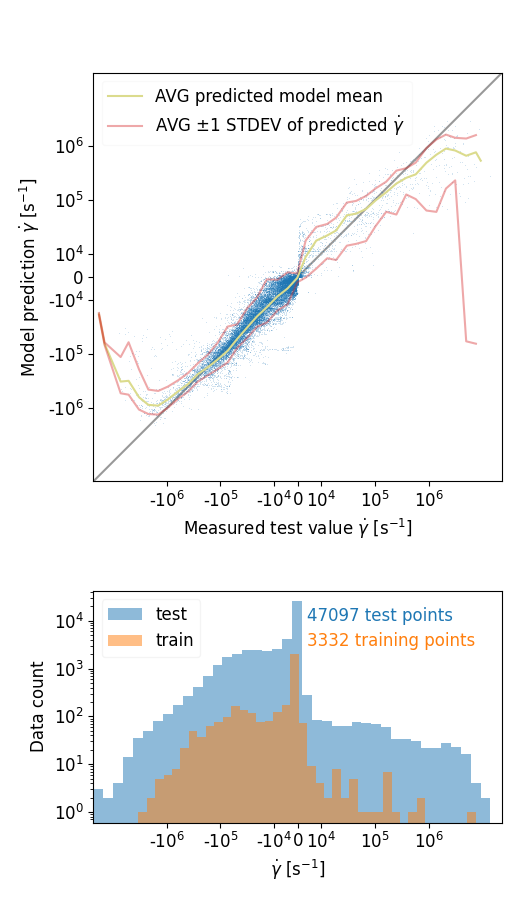}
\includegraphics[width=0.2\textwidth]{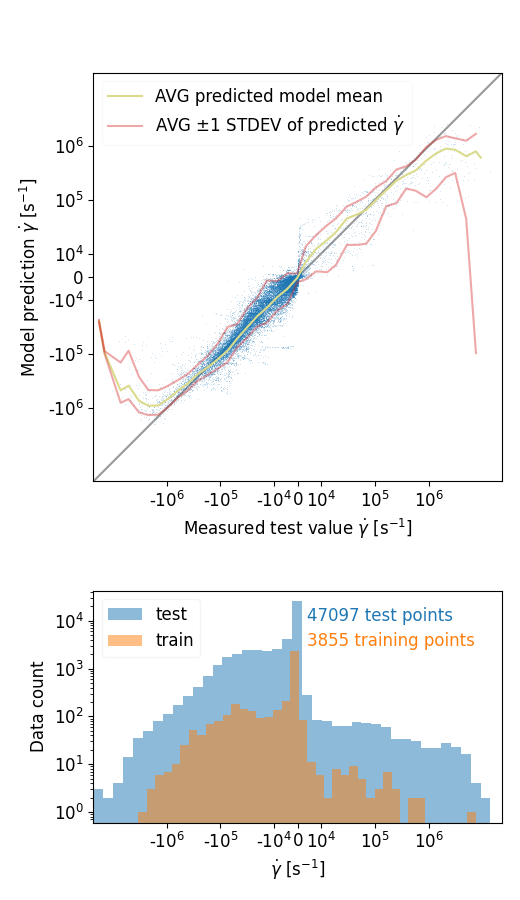}
\includegraphics[width=0.2\textwidth]{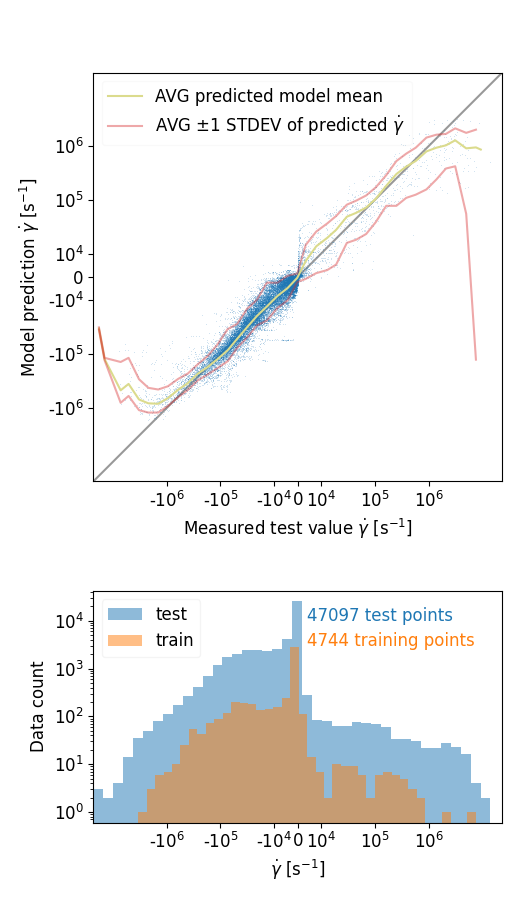}
\includegraphics[width=0.2\textwidth]{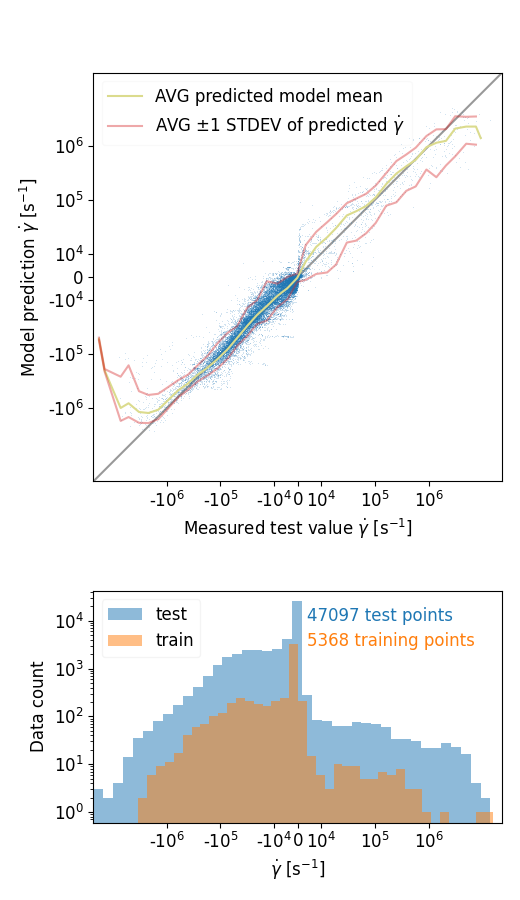}
\includegraphics[width=0.2\textwidth]{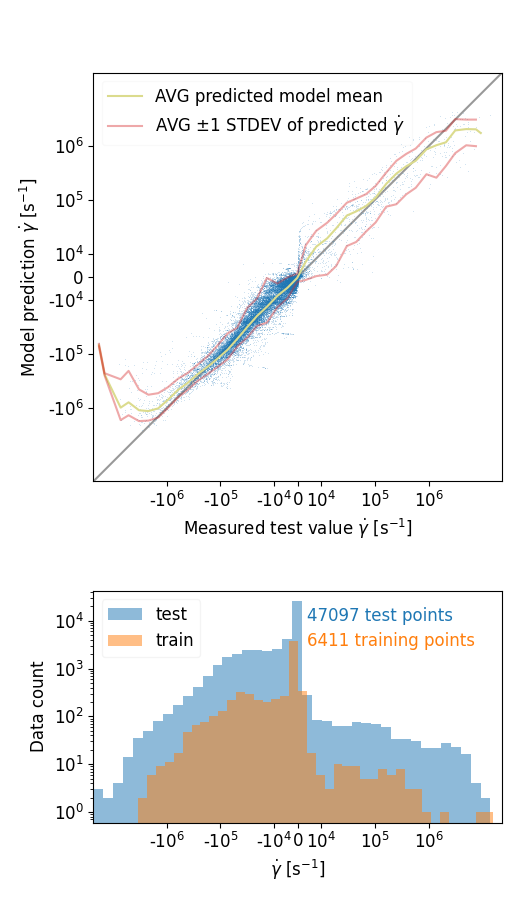}
\includegraphics[width=0.2\textwidth]{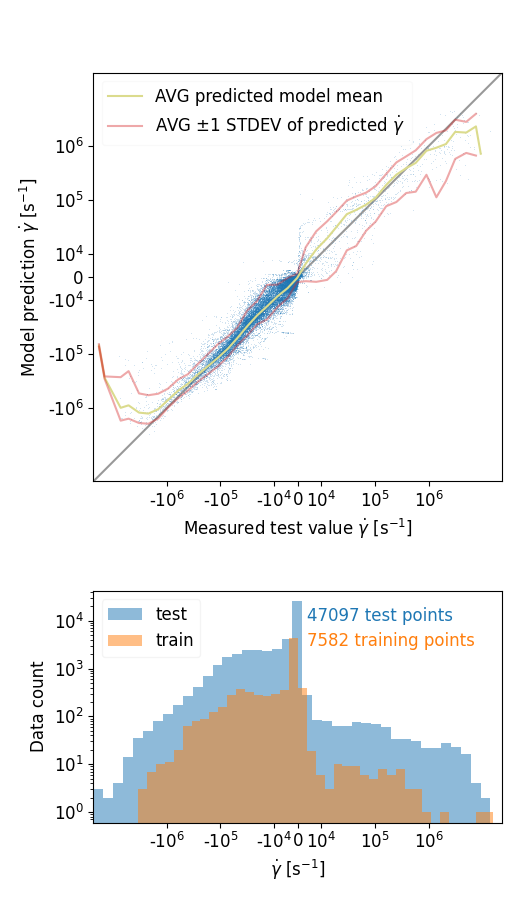}
\includegraphics[width=0.2\textwidth]{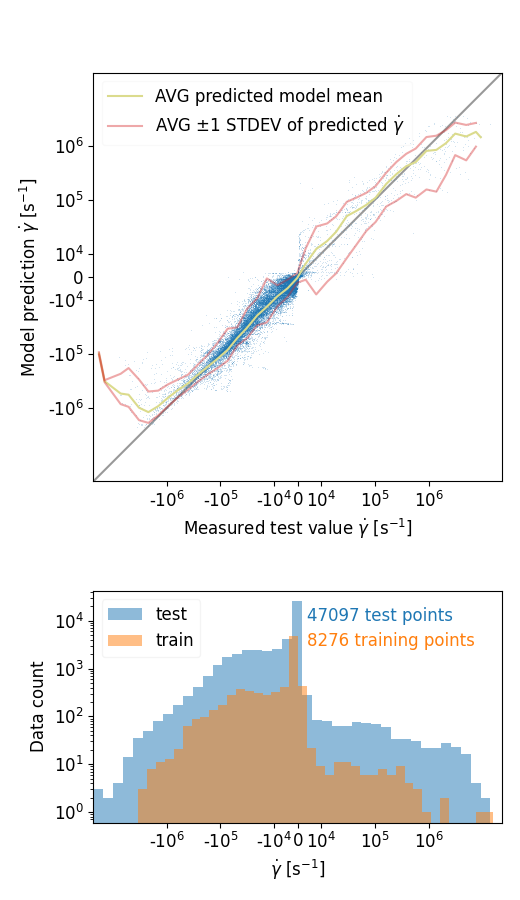}
\includegraphics[width=0.2\textwidth]{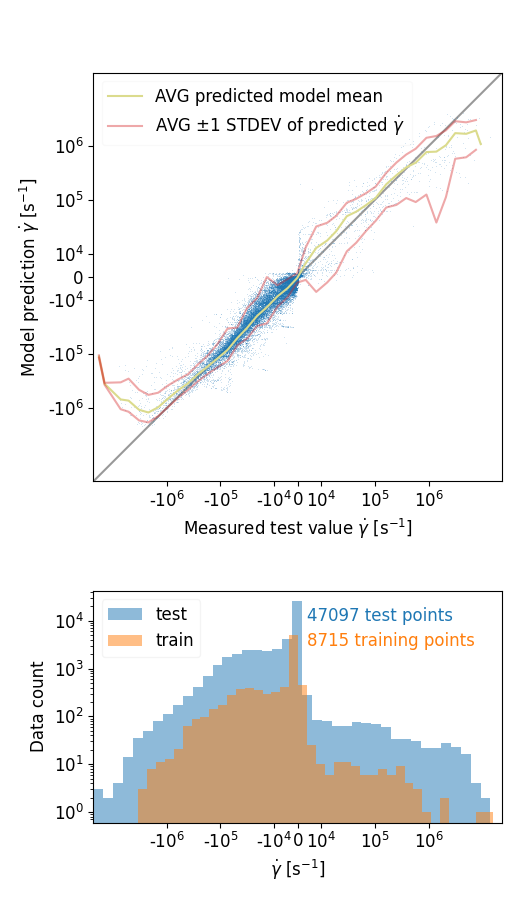}
\includegraphics[width=0.2\textwidth]{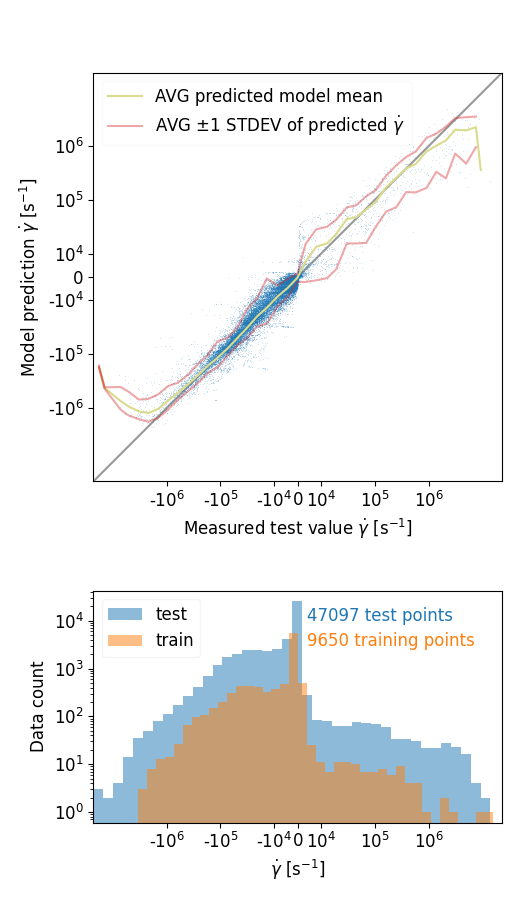}
   \end{center}
   \caption{Parity plots representing evolution of the model quality from the bootstrap data set to the fifteenth iteration of the BO sequence, ordered left-to-right, top-to-bottom.}
   \label{fig_parity_plot_evolution}
\end{figure*}

Evolution of the parity plots and distribution of strain rates in the training set is shown in Fig.~\ref{fig_parity_plot_evolution}.
Similar to the model error metrics plotted in Fig.~\ref{fig_convergence_plots}, the most drastic improvements occurred in the first few iterations of the sequence,
but in these plots we can see the gradual refinement of posterior mean and uncertainty over the entire range of strain rate values.

\section{Slip system parameters}\label{secA2}
\begin{figure}
    \begin{center}
      \includegraphics[width=0.35\textwidth]{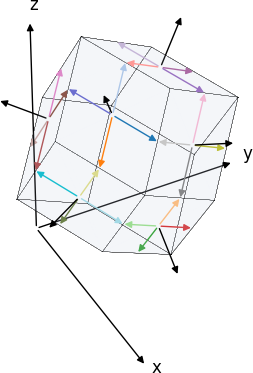}
   \caption{Orientation of 
   $\langle$111$\rangle$\{110\}
   BCC slip systems within the simulation box whose edges are parallel to the coordinate system labeled x, y, and z. Burgers vectors are drawn (colored arrows) within their slip planes surrounding the corresponding plane normals (black arrows), forming a rhombic dodecahedron. This rotation of the crystal by 30 degrees about the $[\bar{1}11]$ vector is chosen to reduce self interaction of dislocations due to periodic boundary conditions.
   }\label{rhombic_dodecahedron}
   \end{center}
\end{figure}

Table~\ref{table_ss_enum} shows the explicit enumeration of slip system Burgers' vectors and plane normals used in this work and illustrated in Figure~\ref{rhombic_dodecahedron}.
      \begin{table}
         \begin{tabular}{|c|p{2.2cm}|p{2.2cm}|}
            \hline
            i & $\vec{b}_{i}$ & $\hat{n}_{i}$\\
            \hline
            1 & $\frac{b}{\sqrt{3}}[1, 1, -1]$&$\frac{1}{\sqrt{2}}(1, 0, 1)$\\
            2 & $\frac{b}{\sqrt{3}}[-1, -1, 1]$&$\frac{1}{\sqrt{2}}(1, 0, 1)$\\
            3 & $\frac{b}{\sqrt{3}}[-1, 1, 1]$&$\frac{1}{\sqrt{2}}(-1, -0, -1)$\\
            4 & $\frac{b}{\sqrt{3}}[1, -1, -1]$&$\frac{1}{\sqrt{2}}(-1, -0, -1)$\\
            5 & $\frac{b}{\sqrt{3}}[1, 1, 1]$&$\frac{1}{\sqrt{2}}(-1, -0, 1)$\\
            6 & $\frac{b}{\sqrt{3}}[-1, -1, -1]$&$\frac{1}{\sqrt{2}}(-1, -0, 1)$\\
            7 & $\frac{b}{\sqrt{3}}[1, -1, 1]$&$\frac{1}{\sqrt{2}}(1, 0, -1)$\\
            8 & $\frac{b}{\sqrt{3}}[-1, 1, -1]$&$\frac{1}{\sqrt{2}}(1, 0, -1)$\\
            9 & $\frac{b}{\sqrt{3}}[1, -1, 1]$&$\frac{1}{\sqrt{2}}(0, 1, 1)$\\
            10 & $\frac{b}{\sqrt{3}}[-1, 1, -1]$&$\frac{1}{\sqrt{2}}(0, 1, 1)$\\
            11 & $\frac{b}{\sqrt{3}}[1, 1, -1]$&$\frac{1}{\sqrt{2}}(-0, -1, -1)$\\
            12 & $\frac{b}{\sqrt{3}}[-1, -1, 1]$&$\frac{1}{\sqrt{2}}(-0, -1, -1)$\\
            13 & $\frac{b}{\sqrt{3}}[1, 1, 1]$&$\frac{1}{\sqrt{2}}(-0, 1, -1)$\\
            14 & $\frac{b}{\sqrt{3}}[-1, -1, -1]$&$\frac{1}{\sqrt{2}}(-0, 1, -1)$\\
            15 & $\frac{b}{\sqrt{3}}[-1, 1, 1]$&$\frac{1}{\sqrt{2}}(0, -1, 1)$\\
            16 & $\frac{b}{\sqrt{3}}[1, -1, -1]$&$\frac{1}{\sqrt{2}}(0, -1, 1)$\\
            17 & $\frac{b}{\sqrt{3}}[-1, 1, 1]$&$\frac{1}{\sqrt{2}}(1, 1, 0)$\\
            18 & $\frac{b}{\sqrt{3}}[1, -1, -1]$&$\frac{1}{\sqrt{2}}(1, 1, 0)$\\
            19 & $\frac{b}{\sqrt{3}}[1, -1, 1]$&$\frac{1}{\sqrt{2}}(-1, -1, -0)$\\
            20 & $\frac{b}{\sqrt{3}}[-1, 1, -1]$&$\frac{1}{\sqrt{2}}(-1, -1, -0)$\\
            21 & $\frac{b}{\sqrt{3}}[1, 1, 1]$&$\frac{1}{\sqrt{2}}(1, -1, -0)$\\
            22 & $\frac{b}{\sqrt{3}}[-1, -1, -1]$&$\frac{1}{\sqrt{2}}(1, -1, -0)$\\
            23 & $\frac{b}{\sqrt{3}}[1, 1, -1]$&$\frac{1}{\sqrt{2}}(-1, 1, 0)$\\
            24 & $\frac{b}{\sqrt{3}}[-1, -1, 1]$&$\frac{1}{\sqrt{2}}(-1, 1, 0)$\\
            \hline
         \end{tabular}
         \caption{Enumeration of slip systems}\label{table_ss_enum}
      \end{table}
\end{appendices}
\end{document}